\documentclass[prc,nofootinbib,twocolumn]{revtex4}
\usepackage{amsmath}
\usepackage{amssymb}
\usepackage{longtable}
\usepackage{graphicx}
\usepackage{fleqn}
\usepackage{epsfig}
\usepackage[figuresright]{rotating}

\input{tcilatex}

\begin{document}

\preprint{}
\author{P. E. Koehler }
\email{p.e.koehler@fys.uio.no}
\affiliation{Department of Physics, University of Oslo, N-0316 Oslo, Norway}
\author{K. H. Guber}
\affiliation{Reactor and Nuclear Systems Division, Oak Ridge National Laboratory,Oak
Ridge, Tennessee 37831, USA}
\title{Improved $^{192,194,195,196}$Pt($n,\gamma $) and $^{192}$Ir($n,\gamma 
$) astrophysical reactions rates}
\date{\today }

\begin{abstract}
$^{192}$Pt is produced solely by the slow neutron capture (\textit{s})
nucleosynthesis process and hence an accurate ($n$,$\gamma $) reaction rate
for this nuclide would allow its use as an important calibration point near
the termination of the \textit{s}-process nucleosynthesis flow. For this
reason, we have measured neutron capture and total cross sections for $%
^{192,194,195,196,nat}$Pt in the energy range from 10 eV to several hundred
keV at the Oak Ridge Electron Linear Accelerator. Measurements on the other
Pt isotopes were, in part, necessitated by the fact that only a relatively
small $^{192}$Pt sample of modest enrichment was available. Astrophysical $%
^{192,194,195,196}$Pt($n,\gamma $) reaction rates, accurate to approximately
3\%--5 \%, were calculated from these data. No accurate reaction rates have
been published previously for any of these isotopes. At \textit{s}-process
temperatures, previously recommended rates are larger (by as much as 35\%)
and have significantly different shapes as functions of temperature, than
our new rates. We used our new Pt results, together with $^{191,193}$Ir(n,$%
\gamma $) data, to calibrate nuclear statistical model calculations and
hence obtain an improved rate for the unmeasured \textit{s}-process
branching-point isotope $^{192}$Ir.
\end{abstract}

\maketitle

\section{Introduction}

\begin{figure}[tbp]
\includegraphics*[width=85mm,keepaspectratio]{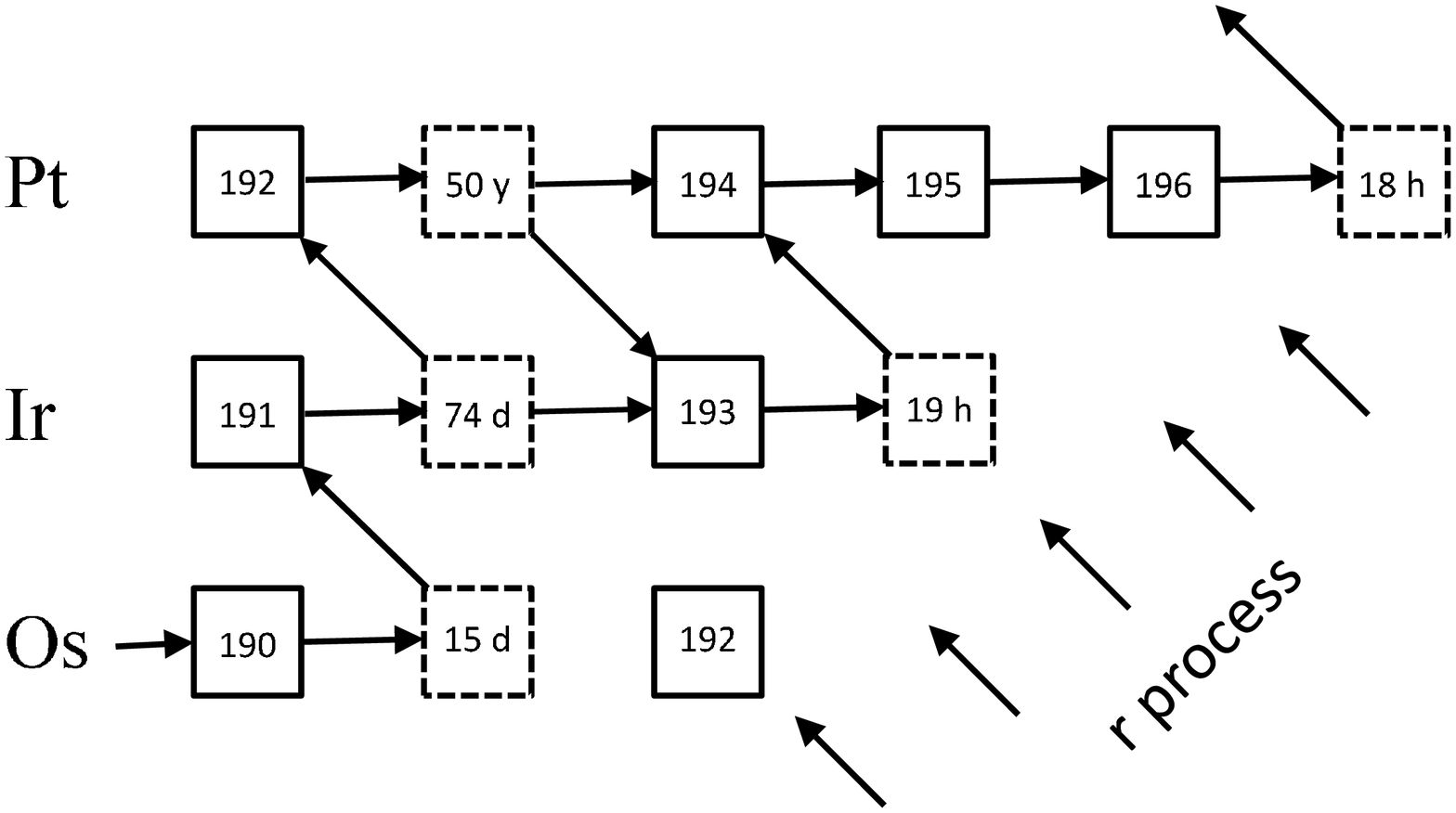}%
\caption{Schematic diagram of the \textit{s}-process path near Pt. Solid and
dashed boxes represent stable and radioactive isotopes, respectively. Stable
isotopes are labeled by their mass numbers whereas half lives serve as
labels for radioactive ones. Contributions to this mass region due to $%
\protect\beta $ decays following the \textit{r} process, whose path runs
through more neutron-rich nuclides, are represented by diagonal arrows.
Although $^{192}$Pt is partially bypassed during the \textit{s} process by a
branching at $^{192}$Ir, it is shielded against contributions from the 
\textit{r} process by its stable isobar $^{192}$Os; hence, $^{192}$Pt is an 
\textit{s}-only isotope.}
\label{PtsProcessPath}
\end{figure}

The $^{192}$Pt($n$,$\gamma $) cross section is particularly important to
nuclear astrophysics studies for at least two reasons, as illustrated in
Fig. \ref{PtsProcessPath}. First, $^{192}$Pt is the heaviest nuclide
produced solely by the \textit{s} process (the so-called \textit{s}-only
isotopes) for which reliable neutron capture data do not exist across the
range of temperatures ($kT=$ 6--26 keV) needed by \textit{s}-process models.
Second, $^{192}$Pt is partially bypassed during the \textit{s} process by a
branching at $^{192}$Ir. Because this branching is expected to be
practically independent of temperature and electron density, it is
particularly sensitive to the neutron density during the \textit{s} process.
However, a meaningful analysis of this branching has not been possible
because of the large uncertainty in the $^{192}$Pt($n$,$\gamma $) reaction
rate.

The only previously reported measurement \cite{Ne92} of the $^{192}$Pt($n$,$%
\gamma $) reaction rate employed an activation technique, so the reaction
rate was obtained only at one temperature ($kT=25$ keV) and with a fairly
large uncertainty (29\%). Activation measurements are hampered by the
comparatively long half-life of $^{193}$Pt, and by the facts that both the
half life and decay intensities for this isotope are fairly uncertain.
Theoretical estimates of reaction rates for the Pt isotopes \cite%
{Ha81,Ho76,Ra98,Go98,Ra2000a}, using statistical models, differ by up to a
factor of 3. The currently recommended $^{192}$Pt($n$,$\gamma $) rate \cite%
{Ba2000,Di2006,Di2009}, which is based on a semi-empirical estimate, is a
factor of 3 larger than that obtained in the activation measurement.

An experimental determination of the $^{192}$Pt($n$,$\gamma $) reaction rate
using a time-of-flight technique has been stymied by the very small (0.79\%)
natural abundance of this nuclide. The necessary gram-size, high enrichment
sample has not been available. However, given improvements made in the
neutron capture apparatus at the Oak Ridge Electron Linear Accelerator
(ORELA) (e.g. see Ref. \cite{Ko2000}), measurements with the available,
relatively small (700 mg of Pt), sample of modest enrichment (56.7\%) seemed
feasible. The small sample size required a fairly long measurement time and
the use of the same material for both the neutron capture and transmission
measurements (with the shape of the sample refabricated between
measurements). The low enrichment required complementary measurements on $%
^{194,195,196}$Pt.

One benefit of having to make all of these measurements is that these data
should be very useful for testing and improving the nuclear statistical
model, which still must be relied upon to calculate reaction rates for
nuclides such as $^{192}$Ir that are beyond the reach of current measurement
techniques. Hence, we used our new data, together with exisiting $^{191,193}$%
Ir(n,$\gamma $) cross sections \cite{Ma78} to constrain statistical-model
parameters and obtain an improved prediction for the $^{192}$Ir(n,$\gamma $)
rate at \textit{s}-process temperatures.

\section{Experiment and data reduction\label{ExpDetailsSection}}

The experimental apparatus has been described previously many times (e.g.,
see Ref \cite{Ko2001} and references contained therein), so only the salient
features will be mentioned herein. The ORELA \cite{Pe82,Bo90,Gu97b} was
operated at a pulse rate of 525 Hz, a pulse width of 8 ns, and a power of
7-8 kW. Neutron energy was determined by time of flight. The samples were
isotopically enriched metallic platinum. The isotopic compositions and
sample thicknesses are given in Table \ref{SampTable}. With the exception of
the $^{192}$Pt and natural samples, enrichments were greater than 96\%.

\begin{table*}[tbp] \centering%
\caption{Isotopic compositions and thicknesses of samples.\label{SampTable}} 
\begin{tabular}{ccccccccc}
\hline\hline
Sample & \multicolumn{6}{c}{Atomic percent} & \multicolumn{2}{c}{Sample
thickness (10$^{-3}$ atom/b)} \\ 
& $^{190}$Pt & $^{192}$Pt & $^{194}$Pt & $^{195}$Pt & $^{196}$Pt & $^{198}$Pt
& Capture & Transmission \\ \hline
$^{192}$Pt & \TEXTsymbol{<}0.5 & 56.97 & 26.16 & 11.23 & 4.70 & 0.90 & 0.463
& 4.53 \\ 
$^{194}$Pt & \TEXTsymbol{<}0.01 & 0.026 & 96.46 & 2.44 & 0.90 & 0.16 & 1.68
& 23.6 \\ 
$^{195}$Pt & 0.04 & 0.01 & 1.21 & 97.29 & 1.40 & 0.10 & 0.607 & 23.8 \\ 
$^{196}$Pt & \TEXTsymbol{<}0.01 & 0.04 & 0.67 & 1.69 & 97.25 & 0.28 & 1.71 & 
14.4 \\ 
Natural & 0.01 & 0.79 & 32.9 & 33.8 & 25.3 & 7.20 & - & 7.57 \\ \hline\hline
\end{tabular}
\end{table*}%

Neutron capture measurements were made on flight path 7 at a
source-to-sample distance of 40.12 m using a pair of C$_{6}$D$_{6}$
detectors, and employed the pulse-height-weighting technique. A $^{10}$B
filter was used to remove overlap neutrons from preceding beam bursts, and a
Pb filter was used to reduce $\gamma $-flash effects. These filters were
placed in the beam at a distance of 5 m from the neutron source. Cross
section normalization was made via the saturated resonance technique \cite%
{Ma79}\ using the 4.9-eV resonance in the $^{197}$Au($n$,$\gamma $) reaction.

A thin $^{6}$Li-loaded glass scintillator \cite{Ma71a}, located 43 cm ahead
of the sample in the neutron beam, was used to measure the energy dependence
of the neutron flux. Separate sample-out background measurements were made,
and measurements with a carbon sample were used to subtract the very small,
smoothly varying background caused by sample-scattered neutrons.

Total neutron cross sections were measured via transmission on flight path 1
using a $^{6}$Li-loaded glass scintillator at a source-to-detector distance
of 79.827 m. The measurements were made at the same time, and hence under
the same ORELA operating conditions, as the (\textit{n},$\gamma $)
experiments. The samples were cylindrical in shape, being between 7.9 to
16.8 mm in diameter. A $^{10}$B filter was used to remove overlap neutrons
from preceding beam bursts, and a Pb filter was used to reduce $\gamma $%
-flash effects. These filters were placed in the beam at a distance of 5 m
from the neutron source. Separate runs for each sample were made at a pulse
rate of 130 Hz to determine the residual background due to overlap neutrons
from preceding beam bursts. These runs were made at the same time as the $%
^{197}$Au($n$,$\gamma $) calibration measurements on flight path 7. The Pt
samples were exchanged periodically with an empty sample holder, and with
polyethylene and bismuth absorbers, which were used for determination of
backgrounds.

$\mathcal{R}$-matrix analysis of the resolved resonance region will be
described in a forthcoming paper. In the unresolved resonance region, the
capture data were averaged over coarse energy bins and the relatively small
corrections for multiple scattering and resonance self-shielding were
calculated using the code \textsc{SESH} \cite{Fr68}. These data also were
corrected for isotopic impurities in the samples using the current
measurements.

\section{Average cross sections and astrophysical reaction rates\label%
{AvgSigAndRRSection}}

\begin{figure}[tbp]
\includegraphics*[width=85mm,keepaspectratio]{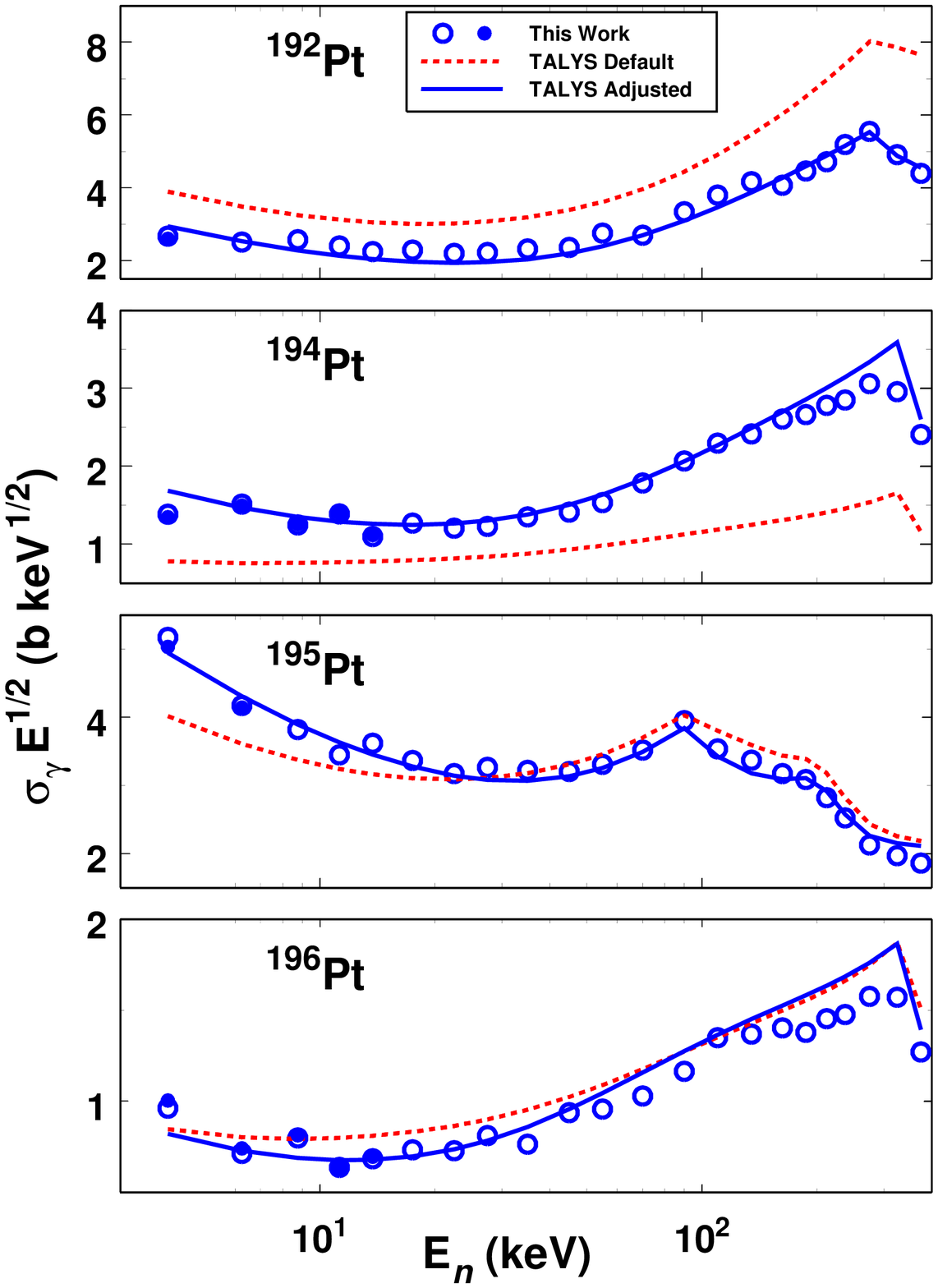}%
\caption{(Color online) Reduced cross sections for $^{192,194,195,196}$Pt(%
\textit{n},$\protect\gamma $) averaged over coarse bins from 3 to 400 keV.
Open symbols depict results obtained directly from the time-of-flight data
whereas filled symbols show averaged cross sections calculated from the
resonance parameters. Error bars, representing one-standard-deviation
statistical uncertainties, are smaller than the symbols. Dashed red curves
depict predictions, using default parameters, of the statistical model code 
\textsc{TALYS} \protect\cite{Ko2008}. Solid blue curves depict \textsc{TALYS}
calculations after calibration of model parameters using average resonance
parameters. See text for details.}
\label{PtUnResCapFig}
\end{figure}

Cross sections averaged over coarse energy bins typically used in the
calculation of astrophysical reaction rates are shown in Table \ref%
{PtUnResCapTable} and Fig. \ref{PtUnResCapFig}. The cross sections in this
figure have been multiplied by the square root of the energy at the center
of each bin, effectively removing the underlying $1/v$ component, so that
data over a wide range of energies for all four isotopes can be shown in the
same graph. Open symbols in this figure represent average cross sections
calculated directly from numerical integration of the data. Corrections for
multiple scattering, self shielding, and isotopic impurities were calculated
as described above. Solid symbols in Fig. \ref{PtUnResCapFig} represent
average cross sections calculated from the resonance parameters. Of course,
in this case, corrections to the data are calculated on a
resonance-by-resonance basis. Uncertainties common to both methods of
calculating average cross section (e.g., due to normalization) are not
included in this table or figure, and therefore represent
one-standard-deviation statistical uncertainties only. The good agreement
between average cross sections obtained by the two techniques attests to the
accuracy of the background subtraction and corrections we applied to our
data. All four data sets show effects of neutron inelastic channels opening
up, near 100 keV in $^{195}$Pt and 300 keV in $^{192,194,196}$Pt.

\begingroup
\squeezetable
\begin{table*}[tbp] \centering%
\caption{Cross sections averaged over coarse energy bins  for the
$^{192,194,195,195}$Pt($n,\gamma$) reactions.\label{PtUnResCapTable}} 
\begin{tabular}{ccccccccc}
\hline\hline
Energy Range & \multicolumn{8}{c}{Neutron Capture Cross Section (mb)} \\ 
\cline{2-9}
(keV) & \multicolumn{2}{c}{$^{192}$Pt} & \multicolumn{2}{c}{$^{194}$Pt} & 
\multicolumn{2}{c}{$^{195}$Pt} & \multicolumn{2}{c}{$^{196}$Pt} \\ 
\cline{2-9}
& Numerical & Resonance & Numerical & Resonance & Numerical & Resonance & 
Numerical & Resonance \\ 
& Integration &  & Integration &  & Integration &  & Integration &  \\ \hline
3-5 & $1339\pm 30$ & $1302\pm 29$ & $694.3\pm 4.8$ & $674.8\pm 4.7$ & $%
2586\pm 22$ & $2515\pm 21$ & $481.2\pm 3.9$ & $502.2\pm 4.1$ \\ 
5-7.5 & $1004\pm 22$ &  & $607.0\pm 4.6$ & $598.3\pm 4.5$ & $1668\pm 20$ & $%
1657\pm 20$ & $285.4\pm 2.9$ & $296.6\pm 3.0$ \\ 
7.5-10 & $871\pm 22$ &  & $422.6\pm 4.7$ & $426.1\pm 4.7$ & $1292\pm 10$ & 
& $270.1\pm 3.2$ & $275.1\pm 3.3$ \\ 
10-12.5 & $718\pm 22$ &  & $414.4\pm 4.6$ & $416.3\pm 4.6$ & $1028\pm 10$ & 
& $189.8\pm 2.6$ & $189.8\pm 2.6$ \\ 
12.5-15 & $606\pm 22$ &  & $296.7\pm 3.8$ & $301.9\pm 3.9$ & $976\pm 10$ & 
& $184.5\pm 3.0$ & $187.6\pm 3.0$ \\ 
15-20 & $547\pm 15$ &  & $304.1\pm 2.8$ &  & $804.4\pm 7.4$ &  & $175.3\pm
1.9$ &  \\ 
20-25 & $464\pm 15$ &  & $254.5\pm 2.7$ &  & $669.3\pm 7.1$ &  & $153.7\pm
1.9$ &  \\ 
25-30 & $424\pm 15$ &  & $234.8\pm 2.8$ &  & $622.8\pm 7.5$ &  & $155.0\pm
2.0$ &  \\ 
30-40 & $393\pm 11$ &  & $228.8\pm 2.0$ &  & $544.7\pm 5.5$ &  & $129.4\pm
1.4$ &  \\ 
40-50 & $352\pm 12$ &  & $210.4\pm 2.3$ &  & $477.9\pm 5.9$ &  & $139.8\pm
1.7$ &  \\ 
50-60 & $372\pm 12$ &  & $207.2\pm 2.2$ &  & $446.2\pm 5.5$ &  & $129.1\pm
1.6$ &  \\ 
60-80 & $322.7\pm 9.6$ &  & $213.8\pm 1.9$ &  & $420.5\pm 4.5$ &  & $%
123.1\pm 1.3$ &  \\ 
80-100 & $352.7\pm 8.9$ &  & $218.3\pm 1.8$ &  & $416.5\pm 4.2$ &  & $%
122.9\pm 1.3$ &  \\ 
100-120 & $362.8\pm 9.5$ &  & $219.1\pm 1.9$ &  & $337.3\pm 4.2$ &  & $%
128.7\pm 1.4$ &  \\ 
120-150 & $359.2\pm 8.0$ &  & $208.1\pm 1.6$ &  & $290.3\pm 3.4$ &  & $%
117.9\pm 1.1$ &  \\ 
150-175 & $319.6\pm 8.4$ &  & $204.6\pm 1.7$ &  & $248.9\pm 3.6$ &  & $%
110.1\pm 1.2$ &  \\ 
175-200 & $325.8\pm 8.3$ &  & $194.6\pm 1.8$ &  & $225.4\pm 3.6$ &  & $%
100.8\pm 1.2$ &  \\ 
200-225 & $323.9\pm 8.9$ &  & $190.9\pm 1.8$ &  & $193.6\pm 3.5$ &  & $%
99.7\pm 1.2$ &  \\ 
225-250 & $337.1\pm 8.8$ &  & $184.9\pm 1.8$ &  & $164.0\pm 3.4$ &  & $%
95.9\pm 1.2$ &  \\ 
250-300 & $334.8\pm 6.6$ &  & $184.6\pm 1.3$ &  & $128.3\pm 2.3$ &  & $%
95.13\pm 0.90$ &  \\ 
300-350 & $272.8\pm 5.8$ &  & $164.1\pm 1.2$ &  & $109.4\pm 2.1$ &  & $%
87.22\pm 0.85$ &  \\ 
350-400 & $227.1\pm 5.9$ &  & $124.5\pm 1.2$ &  & $96.1\pm 2.1$ &  & $%
65.65\pm 0.78$ &  \\ 
400-450 & $213.1\pm 5.7$ &  & $109.1\pm 1.1$ &  & $87.6\pm 2.1$ &  & $%
52.45\pm 0.70$ &  \\ 
450-500 & $214.5\pm 5.5$ &  & $102.4\pm 1.1$ &  & $79.5\pm 2.1$ &  & $%
48.59\pm 0.65$ &  \\ \hline\hline
\end{tabular}
\end{table*}
\endgroup%

\textsc{SAMMY} was used to calculate astrophysical reaction rates from the
resonance parameters together with the averaged cross sections shown in Fig. %
\ref{PtUnResCapFig}. Resulting Maxwellian-averaged cross sections are given
in Table \ref{RRTable}. As a check on the \textsc{SAMMY} calculations,
reaction rates also were calculated following the technique of Ref. \cite%
{Be92}. The two methods agree to within 0.5\%. Statistical uncertainties in
the reaction rates are negligible when compared to the overall normalization
uncertainty. From the uncertainties in the $^{197}$Au(\textit{n},$\gamma $)
and $^{6}$Li(\textit{n},$\alpha $) cross sections, the statistical precision
of the calibration measurements, the repeatability of the calibration runs,
and uncertainties in the sample sizes and isotopics, we calculated
normalization uncertainties of 3\%--4 \% in the reaction rates.

\begin{table*}[tbp] \centering%
\caption{Astrophysical rates for the $^{192,194,195,195}$Pt($n,\gamma$)
reactions.\label{RRTable}} 
\begin{tabular}{ccccc}
\hline\hline
Thermal energy $kT$ & \multicolumn{4}{c}{$\langle \sigma v\rangle /v_{T}$
(mb)} \\ 
(keV) & $^{192}$Pt & $^{194}$Pt & $^{195}$Pt & $^{196}$Pt \\ \hline
5 & $1261\pm 52$ & $686\pm 21$ & $2148\pm 65$ & $408\pm 13$ \\ 
8 & $920\pm 38$ & $508\pm 15$ & $1497\pm 45$ & $300.5\pm 9.4$ \\ 
10 & $801\pm 33$ & $446\pm 13$ & $1273\pm 38$ & $263.7\pm 8.2$ \\ 
15 & $640\pm 26$ & $361\pm 11$ & $967\pm 29$ & $214.8\pm 6.6$ \\ 
20 & $559\pm 23$ & $320.3\pm 9.4$ & $810\pm 24$ & $190.8\pm 5.9$ \\ 
25 & $513\pm 21$ & $297.2\pm 8.8$ & $713\pm 21$ & $176.7\pm 5.5$ \\ 
30 & $483\pm 20$ & $282.5\pm 8.4$ & $644\pm 19$ & $167.4\pm 5.2$ \\ 
\hline\hline
\end{tabular}
\end{table*}%

We know of no measured $^{192,194,195,196}$Pt($n,\gamma $) reaction rates
published in peer-reviewed journals. Our preliminary reaction rates reported
in Ref. \cite{Ko2002} are in agreement with the present rates to within the
uncertainties. Although there have been numerous semiempirical and
theoretical estimates, we know of only three other actual measurements of
reaction rates to which our results can be compared, and only at a single
temperature.

A measurement of the $^{192}$Pt(\textit{n},$\gamma $) Maxwellian-averaged
cross section (MACS) at $kT=30$ keV, $\langle \sigma \rangle _{30}=196\pm 56$
mb, 2.5 times smaller than our measured rate, was reported in Ref. \cite%
{Ne92}. This measurement was made using a pseudo-Maxwellian neutron source
at $kT=25$ keV, which was then extrapolated to 30 keV. Activation
measurements of the $^{192}$Pt(\textit{n},$\gamma $) rate are hampered by
the relatively long and somewhat uncertain half-life and the low energy of
the decay of the product nuclide. Hence, it may not be too surprising that
the most recent reaction-rate compilation \cite{Ba2000} recommends a
semi-empirical estimate ($590\pm 120$ mb \cite{Be97a}) for this rate over
the measurement of Ref. \cite{Ne92}. Note that $^{192,194,195}$Pt($n,\gamma $%
) reaction rates in the more recent KADoNiS compilation \cite{Di2006,Di2009}
are identical to those in Ref. \cite{Ba2000}.

There have been two previously reported \cite{Be97a,Ma2008} measurements of
the $^{196}$Pt($n,\gamma $) reaction rate, both using the same technique as
mentioned in the previous paragraph. Uncertainties in these previous results
are about four times larger than ours. The most recent previous result ($%
171\pm 22$ mb at $kT=30$ keV) \cite{Ma2008} agrees with ours to within the
quoted uncertainties, whereas the older result ($197\pm 23$ mb $kT=30$ keV) 
\cite{Be97a} is just outside the combined uncertainties. The currently
recommended rate \cite{Di2009} is the weighted average of these two results
together with the temperature dependence predicted by the statistical model
calculation of Ref. \cite{Ra2000a}.

\begin{figure}[tbp]
\includegraphics*[width=85mm,keepaspectratio]{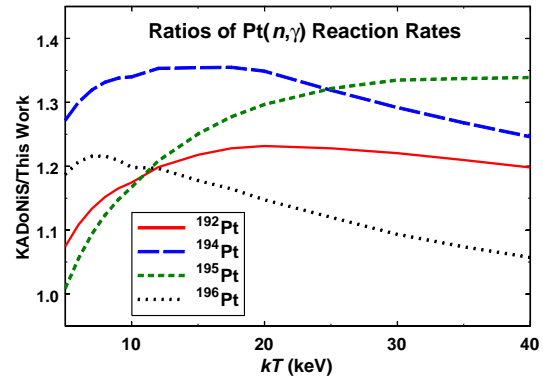}%
\caption{(Color online) Ratios of astrophysical rates from Ref. \protect\cite%
{Di2009} to our new rates for the $^{192,194,195,196}$Pt(\textit{n},$\protect%
\gamma $) reactions, depicted as solid red, long-dashed blue, short-dashed
green, and dotted black curves, respectively.}
\label{PtAllRRKADdUs}
\end{figure}

Our new $^{192,194,195,196}$Pt(\textit{n},$\gamma $) reaction rates are
compared to recommended rates from the most recent compilation \cite%
{Di2006,Di2009} in Fig. \ref{PtAllRRKADdUs}. At \textit{s}-process
temperatures, our rates are far more accurate (uncertainties reduced by
factors of 4 to 8) than the previously recommended rates. In general,
previously recommended rates are larger (by as much as 35\% at \textit{s}%
-process temperatures) and have significantly different shapes, as functions
of temperature, than our new rates.

\section{\textit{s}-process neutron density and improved $^{192}$Ir(\textit{n%
},$\protect\gamma $) reaction rate}

Although $^{192}$Pt is an \textit{s}-only isotope, it is partially bypassed
during the \textit{s} process by a branching at $^{192}$Ir. Because this
branching is expected to be practically independent of temperature and
electron density, it is particularly sensitive to the neutron density during
the \textit{s} process. A classical branching analysis \cite{Ko2002} using
our preliminary reaction rates resulted in a significantly lower neutron
density [$n_{n}=(7_{-2}^{+5})\times 10^{7}$ cm$^{-3}$], inconsistent with
the density [$n_{n}=(4.1\pm 0.6)\times 10^{8}$ cm$^{-3}$] resulting from
analyses \cite{To95} of several other branchings. Because our new rates are
consistent with our preliminary ones to within the uncertainties, we confirm
this result. In contrast, just the opposite conclusion [$%
n_{n}=(4.3_{-2.5}^{+3.4})\times 10^{8}$ cm$^{-3}$] was reached in Ref. \cite%
{Ne92}, based on a classical branching analysis using the only previous
measurement of the $^{192}$Pt($n,\gamma $) reaction rate. The large
reduction in extracted neutron density and its uncertainty from our new
analysis are directly attributable to the substantially larger $^{192}$Pt($%
n,\gamma $) reaction rate and substantially reduced uncertainty,
respectively, from our new measurements.

The large uncertainty in the above estimate of the neutron density from the $%
^{192}$Ir branching is dominated by the assumed \cite{Ko2002} factor of 2
uncertainty in the unmeasured $^{192}$Ir($n,\gamma $) reaction rate.
Theoretical rates (which are based on nuclear statistical model
calculations) for this reaction listed in the latest compilation \cite%
{Di2009} vary by a factor of 2.7. In an attempt to obtain an improved rate
for this reaction, we have used our new Pt data to constrain parameters in
the nuclear statistical model. The code \textsc{TALYS} \cite{Ko2008} was
used for these calculations.

As shown in Fig. \ref{PtUnResCapFig}, average cross sections predicted by 
\textsc{TALYS} \cite{Ko2008} using default parameters require normalizations
ranging from 0.72 to 1.5 to yield reasonable agreement with our data. Also,
the predicted cross section is significantly flatter as a function of energy
than the measured one for $^{195}$Pt($n,\gamma $). As shown in Fig. \ref%
{IrUnResCapFig}, shapes of default \textsc{TALYS} cross sections also are
significantly flatter than the $^{191,193}$Ir(\textit{n},$\gamma $) data of
Ref. \cite{Ma78}. There are substantial disagreements among the various $%
^{191,193}$Ir(\textit{n},$\gamma $) measurements \cite%
{Do67,Li76,Ma78,Ko94a,Ja97,Ta2008}. We have chosen to use the data of Ref. 
\cite{Ma78} because they are the only sets to span the energy range needed
to determine the reaction rate at \textit{s}-process temperatures, and
because other measurements by the same group with the same apparatus have
shown, in general, to be reliable.

\begin{figure}[tbp]
\includegraphics*[width=85mm,keepaspectratio]{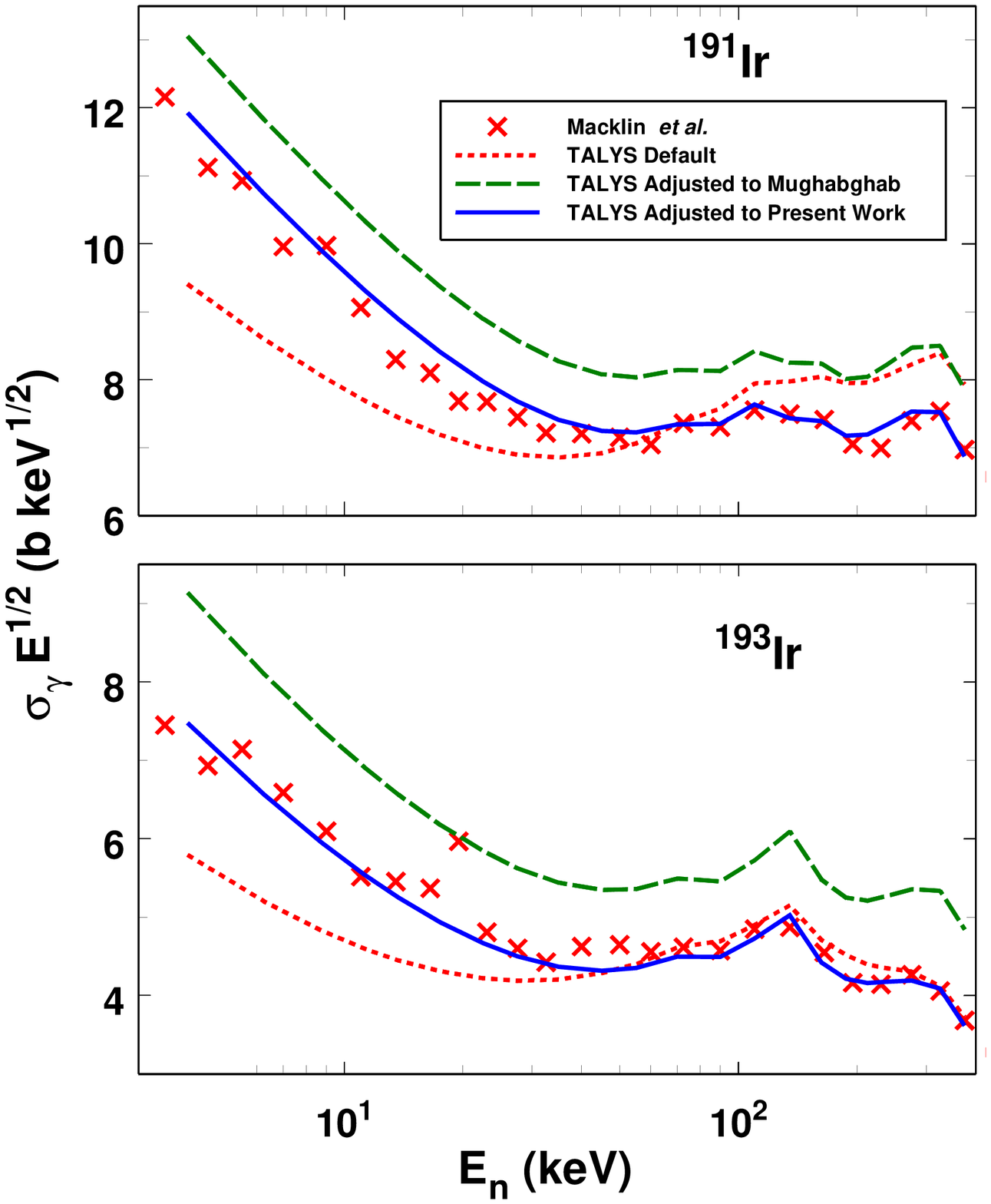}%
\caption{(Color online) Reduced cross sections for $^{191,193}$Ir(\textit{n},%
$\protect\gamma $) averaged over coarse bins from 3 to 400 keV. Data are
those of Macklin \textit{et al}. \protect\cite{Ma78} (X's). Dashed red
curves depict predictions, using default parameters, of the statistical
model code \textsc{TALYS} \protect\cite{Ko2008}. Long-dashed green curves
depict \textsc{TALYS} calculations after adjustment to average resonance
parameters of Ref. \protect\cite{Mu2006}. Solid blue curves depict \textsc{%
TALYS} calculations after adjustment to local systematics. See text for
details.}
\label{IrUnResCapFig}
\end{figure}

There are several model paramenters which can be constrained by measured
average resonance parameters. To this end, we report in Table \ref%
{AveResParTable} $D_{0}$, $S_{0}$, and $<\Gamma _{\gamma 0}>$ values
resulting from $\mathcal{R}$-matrix analysis of the resolved resonance
region. Because \textsc{TALYS} assumes that the Porter-Thomas distribution
(PTD) \cite{Po56} is valid (e.g., in calculating width fluctuation
correction factors), $D_{0}$\ and $S_{0}$ values in Table \ref%
{AveResParTable} were corrected for the effects of missed resonances using
the technique of Ref. \cite{Fu65} and assuming the PTD. The $<\Gamma
_{\gamma 0}>$ values in Tabe \ref{AveResParTable} were determined from a
maximum-likelihood analysis by assuming these widths were $\chi ^{2}$
distributed. The corresponding uncertainties are one-standard-deviation
estimates from the maximum-likelihood analysis. For $^{195}$Pt, separate $%
<\Gamma _{\gamma 0}>$ values were determined for definite 0$^{-}$ and 1$^{-}$
resonance assignments.

\begin{table*}[tbp] \centering%
\caption{Average resonance parameters.\label{AveResParTable}} 
\begin{tabular}{cccc}
\hline\hline
Nuclide & $D_{0}$ (eV) & $10^{4}S_{0}$ & $<\Gamma _{\gamma 0}>$ (meV) \\ 
\hline
$^{192}$Pt & 28.4$\pm 1.2$ & 2.06$\pm 0.23$ & 62.40$\pm 0.95$ \\ 
$^{194}$Pt & 71.8$\pm 2.9$ & 2.01$\pm 0.22$ & 76.7$\pm 1.1$ \\ 
$^{195}$Pt & 15.93$\pm 0.41$ & 1.94$\pm 0.14$ & 109.9$_{-2.7}^{+2.9}$ (0$^{-}
$), 127.3$_{-1.4}^{+1.5}$ (1$^{-}$) \\ 
$^{196}$Pt & 192$\pm 12$ & 1.95$\pm 0.34$ & 85.9$_{-1.9}^{+1.8}$ \\ 
\hline\hline
\end{tabular}
\end{table*}%

To obtain an improved $^{192}$Ir($n,\gamma $) rate, first, we adjusted the
level density (at the neutron separation energy) parameter $a$($S_{n}$) to
obtain agreement with the measured $D_{0}$ values for $^{192,194,195,196}$Pt
(Table \ref{AveResParTable}) and $^{191,192,193}$Ir \cite{Mu2006}. The
resulting $a$($S_{n}$) and $D_{0}$ values are listed in Table \ref%
{TALYSParametersTable}, from which it can be seen that the latter agree with
the measured values in Table \ref{AveResParTable} to within the
uncertainties.

Next, we used the \textquotedblleft gamgam\textquotedblright\ option in 
\textsc{TALYS} to input the measured $<\Gamma _{\gamma 0}>$ values (Table %
\ref{AveResParTable} and Ref. \cite{Mu2006}). These quantities, togeher with
the $D_{0}$ values, are used to normalize the $\gamma $-ray transmission
coefficients in \textsc{TALYS}.

Next, we adjusted parameters of the neutron optical model potential (NOMP)
to obtain agreement with the measured $S_{0}$ values (Table \ref%
{AveResParTable} and Ref. \cite{Mu2006}). The data for $^{192,194,195,196}$%
Pt and $^{191,193}$Ir indicate that $S_{0}=2.0$ to about 10\% or better
accuracy. Therefore, we used this value for $^{192}$Ir, which is consistent
with the much less certain value given in Ref. \cite{Mu2006} ($S_{0}=3.7\pm
1.8$). This step was particularly important for obtaining the correct energy
dependence of the cross sections for $^{195}$Pt and $^{191,193}$Ir, but did
not have much effect for $^{192,194,196}$Pt. In particular, a ratio $%
S_{1}/S_{0}\approx 0.1$ was required for the former cases to obtain
agreement with the cross-section shape versus energy. Sensitivity of the $%
^{195}$Pt and $^{191,193}$Ir($n,\gamma $) cross section shape to the neutron
strength functions appears to be a consequence of the smaller average level
spacings for these nuclides compared to $^{192,194,196}$Pt.

Default $S_{0}$ values in \textsc{TALYS} were about 70\% of the measured
ones. We could obtain both $S_{0}\approx 2\times 10^{-4}$ and $%
S_{1}/S_{0}\approx 0.1$ by adjusting both the $a_{V}$ and $a_{D}$ parameters
in \textsc{TALYS}. Default and adjusted values of varied \textsc{TALYS}
parameters are given in Table \ref{TALYSParametersTable}.

\begin{table*}[tbp] \centering%
\caption{Default and adjusted TALYS parameters for level-density and
neutron-optical-models.\label{TALYSParametersTable}} 
\begin{tabular}{ccccccccccccccc}
\hline\hline
Nuclide & \multicolumn{4}{c}{Default} & \multicolumn{10}{c}{Adjusted} \\ 
& $a$($S_{n}$) (MeV$^{-1}$)\textbf{%
\footnotemark[1]%
} & $D_{0}$ (eV) & $10^{4}S_{0}$ & $10^{4}S_{1}$ & \multicolumn{3}{c}{$a$($%
S_{n}$) (MeV$^{-1}$)} & \multicolumn{3}{c}{$D_{0}$ (eV)} & $a_{V}$\textbf{%
\footnotemark[2]%
} & $a_{D}$\textbf{%
\footnotemark[2]%
} & $10^{4}S_{0}$ & $10^{4}S_{1}$ \\ 
&  &  &  &  & LD1\textbf{%
\footnotemark[3]%
} & LD2\textbf{%
\footnotemark[3]%
} & LD3\textbf{%
\footnotemark[3]%
} & LD1\textbf{%
\footnotemark[3]%
} & LD2\textbf{%
\footnotemark[3]%
} & LD3\textbf{%
\footnotemark[3]%
} &  &  &  &  \\ \hline
$^{191}$Ir & 23.04 & 2.49 & 1.52 & 0.40 & 23.40 & 20.50 & 21.50 & 2.05 & 2.00
& 2.12 & 0.82 & 0.60 & 2.05 & 0.11 \\ 
$^{192}$Ir & 22.22 & 0.67 & 1.47 & 0.40 & 22.65 & 20.20 & 21.05 & 0.52 & 0.39
& 0.47 & 0.82 & 0.60 & 1.96 & 0.11 \\ 
$^{193}$Ir & 21.41 & 7.01 & 1.43 & 0.40 & 21.90 & 19.90 & 20.60 & 5.35 & 5.26
& 5.65 & 0.80 & 0.60 & 2.04 & 0.10 \\ 
$^{192}$Pt & 24.00 & 21.97 & 1.38 & 0.42 & 23.49 & 19.69 & 21.01 & 28.39 & 
28.44 & 28.47 & 0.80 & 0.68 & 2.04 & 0.13 \\ 
$^{194}$Pt & 19.83 & 201.10 & 1.26 & 0.57 & 21.95 & 17.69 & 18.43 & 71.70 & 
71.96 & 71.66 & 0.74 & 0.75 & 2.02 & 0.19 \\ 
$^{195}$Pt & 20.38 & 17.99 & 1.26 & 0.42 & 20.60 & 17.97 & 18.38 & 15.87 & 
15.89 & 15.91 & 0.78 & 0.75 & 2.03 & 0.15 \\ 
$^{196}$Pt & 19.37 & 350.11 & 1.23 & 0.41 & 20.62 & 17.07 & 17.30 & 192.28 & 
191.91 & 192.06 & 0.78 & 0.76 & 1.96 & 0.16 \\ \hline\hline
\end{tabular}
\footnotetext[1]{Default TALYS level-density is model 1, constant temperature plus Fermi gas \cite{Gi65}.}
\footnotetext[2]{Numbers in these columns are factors by which default TALYS parameters were multiplied.}
\footnotetext[3]{LD1 = TALYS level-density model 1, etc.}
\end{table*}%

As shown in Figs. \ref{PtUnResCapFig} and \ref{IrUnResCapFig}, adjusted 
\textsc{TALYS} cross sections are in good agreement with the $%
^{192,194,195,196}$Pt data, but they are about 15\%--25 \% larger than the $%
^{191,193}$Ir data, indicating that the average resonance parameters in Ref. 
\cite{Mu2006} are inconsistent with the average cross section data of Ref. 
\cite{Ma78}. Therefore, it seems unlikely that the $^{192}$Ir($n,\gamma $)
reaction rate calculated using average resonance parameters from Ref. \cite%
{Mu2006} will be reliable.

Of the three needed paramenters ($S_{0}$, $D_{0}$, and $<\Gamma _{\gamma 0}>$%
), $D_{0}$ is likely the most problematical, especially given that there are
relatively few known resonances for the Ir isotopes and because the
recommended $D_{0}$ values in Ref. \cite{Mu2006} appear to have undergone
significant corrections for missed resonances. In constrast, $S_{0}$ and $%
<\Gamma _{\gamma 0}>$ typically are much less sensitve to such effects, and
the recommended values for the Ir isotopes for these two parameters are in
line with expectations based on our Pt data. Therefore, constraining $D_{0}$
for $^{192}$Ir seems to be key for obtaining more reliable predictions of
the $^{192}$Ir($n,\gamma $) reaction rate.

As noted above, $D_{0}$ and $<\Gamma _{\gamma 0}>$ are used to normalize $%
\gamma $-ray transmission coefficients in \textsc{TALYS}. The equation used
is,

\begin{multline}
\frac{2\pi <\Gamma _{\gamma 0}>}{D_{0}}=G_{norm}\sum_{J}\sum_{\Pi
}\sum_{Xl}\sum_{I^{\prime }=|J-l|}^{J+l}\sum_{\Pi ^{\prime }}
\label{TALYS Eq 4.68} \\
\int_{0}^{S_{n}}dE_{\gamma }T_{Xl}(E_{\gamma }) \\
\times \rho (S_{n}-E_{\gamma },I^{\prime },\Pi ^{\prime })f(X,\Pi ^{\prime
},l),
\end{multline}%
\newline
where the $J$ and $\Pi $ sums are over compound nucleus states with spin $J$
and parity $\Pi $ that can be formed by \textit{s}-wave incident neutrons,
and $I^{\prime }$ and $\Pi ^{\prime }$ denote the spin and parity of the
final states that may be reached in the first step of the $\gamma $-ray
cascade. The $\gamma $-ray transmission coefficient for type $X$ (electric
or magnetic) and multipolarity $l$ for $\gamma $-ray energy $E_{\gamma }$ is
denoted by $T_{Xl}(E_{\gamma })$, the level density at excitation energy $%
S_{n}-E_{\gamma }$ is denoted by $\rho (S_{n}-E_{\gamma },I^{\prime },\Pi
^{\prime })$, and $f(X,\Pi ^{\prime },l)$ denotes the usual multipole
selection rules. It is understood that the integral over $dE_{\gamma }$
includes a summation over discrete states at lower excitation energies. The $%
\gamma $-ray transmission coefficient is related to the $\gamma $-ray
strength function $f_{Xl}(E_{\gamma })$ via

\begin{equation}
T_{Xl}(E_{\gamma })=2\pi f_{Xl}(E_{\gamma })E_{\gamma }^{2l+1}.
\label{TALYS Eq 4.56}
\end{equation}%
Finally, $G_{norm}$ is a normalization factor that ensure the equality of
Eq. \ref{TALYS Eq 4.68}, and hence in practice the $\gamma $-ray strength
functions are muliplied by this factor before they enter the nuclear
reaction calculation in \textsc{TALYS}. Hence, it is possible to fit \textsc{%
TALYS} to a measured ($n,\gamma $) cross section by varying $D_{0}$ [by
adjusting the level-density parameter $a$($S_{n}$)]. Futhermore, the $a$($%
S_{n}$) values, adjusted to yield our $D_{0}$ values for Pt (Table \ref%
{TALYSParametersTable}), are reasonably well fitted by a straight line as a
function of mass number. Therefore, it seems likely that a reasonably
reliable $D_{0}$ for $^{192}$Ir could be obtained by using the average $a$($%
S_{n}$) value for $^{191,193}$Ir, after they had been adjusted to yield
agreement with the capture cross sections.

For these reasons, we used \textsc{TALYS} to fit the measured $^{191,193}$Ir(%
$n,\gamma $) cross sections by adjusting the appropriate $a$($S_{n}$) value
and used the average of the two $a$($S_{n}$) values to predict the $^{192}$%
Ir($n,\gamma $) cross section and reaction rate. We did this for all three
level-density models in \textsc{TALYS} which allow $a$($S_{n}$) to be
adjusted, and the values are given in Table \ref{TALYSParametersTable}. The
resulting $D_{0}$ values are about 2.0, 0.46, and 5.4 eV for $^{191,192,193}$%
Ir, respectively. These adjusted $^{191,193}$Ir $D_{0}$ values are
intermediate to default \textsc{TALYS} values and those in the compilation
of Ref. \cite{Mu2006}, whereas the adjusted $D_{0}$ values for $^{192}$Ir
are all smaller than the default and compilation ones.

The three $^{192}$Ir($n,\gamma $) cross sections calculated as described
above agree to within 10\%. Due to the normaliztion through Eq. \ref{TALYS
Eq 4.68}, all four $\gamma $-ray strength function models yielded very
similar predictions for a given level-density model. To further assess the
uncertainty due to the level density, we varied $a$($S_{n}$) for each model
by the average amount that our Pt $a$($S_{n}$) values (as functions of mass
number, for each level-density model) deviated from linearity. This
increased the maximum deviation of the various predicted $^{192}$Ir($%
n,\gamma $) cross sections to 11.6\%. Our recommended rate is the average of
the largest and smallest rates predicted by \textsc{TALYS} following the
above procedure, with base uncertainty calculated from the range of
predictions. Additional uncertainties in this rate can arise from
normalization [via Eq. \ref{TALYS Eq 4.68}] to the measured average
radiation width for $^{192}$Ir \cite{Mu2006}, which can contribute a maximum
uncertainty of 6\%, and normalizations to the $^{191,193}$Ir($n,\gamma $)
cross sections, which can add another 5\% \cite{Ba2000}. Therefore, we
calculate that the overall uncertainty in the predicted $^{192}$Ir($n,\gamma 
$) reaction rate is about 22\%. Our recommended MACS values for this
reaction are given in Table \ref{Ir192MACSTable}.

\begin{table}[tbp] \centering%
\caption{Recommended $^{192}$Ir($n,\gamma$) MACS.\label{Ir192MACSTable}} 
\begin{tabular}{cc}
\hline\hline
$kT$ (keV) & MACS (mb) \\ \hline
5 & $8900\pm 1900$ \\ 
10 & $6000\pm 1300$ \\ 
15 & $4800\pm 1000$ \\ 
20 & $4080\pm 890$ \\ 
25 & $3590\pm 790$ \\ 
30 & $3220\pm 720$ \\ 
40 & $2670\pm 600$ \\ \hline\hline
\end{tabular}
\end{table}%

In stellar models of the main \textit{s}-process, most of the neutron
exposure occurs at temperatures near $kT=8$ keV followed by a smaller
exposure at about 23 keV. At these two temperatures, our recommended $^{192}$%
Ir($n,\gamma $) MACS values are 2.1 and 1.6, respectively, times larger than
the rate \cite{Ba2000} used in most previous \textit{s}-process
calculations. The larger $^{192}$Ir($n,\gamma $) MACS values we recommend
would result in $^{192}$Pt being more strongly bypassed during the \textit{s}
process. On the other hand, our new $^{192}$Pt($n,\gamma $) MACS value is
smaller than the previously recommended rate \cite{Ba2000}, and this should
result in less destruction of $^{192}$Pt during the \textit{s} process. New,
realistic \textit{s}-process calculations are needed to ascertain the net
effect.

The previously recommended $^{192}$Ir($n,\gamma $) rate is based on
calculations made with the statistical model code \textsc{NON-SMOKER} \cite%
{Ra98a,Ra2000a}, which had been normalized to data for nearby nuclides. The
fact that the \textsc{NON-SMOKER} rate has a flatter temperature dependence
suggests, based on our experience with \textsc{TALYS} as discussed above,
that neutron strength functions (via the NOMP) in \textsc{NON-SMOKER} need
to be adjusted. This surmise is strengthened by the facts that \textsc{%
NON-SMOKER} predictions for the $^{195}$Pt and $^{191,193}$Ir($n,\gamma $)
cross sections also are flatter, as functions of energy, than the data,
whereas predictions for $^{192,194,196}$Pt are in better agreement with the
data in this respect. Current consesus seems to be that ($n,\gamma $) cross
sections are rather insensitive to the adopted NOMP (e.g., see Ref. \cite%
{Ut2010} for a recent example). However, as we have demonstrated above,
there are important exceptions.\ Therefore, it may be worthwhile to
reexamine the sensitivity of theoretical rates to the NOMP, especially for
nuclides predicted to have small average level spacings.

\section{Constraints on photon-strength-function models}

Comparison of our $^{192,194,195,196}$Pt($n,\gamma $) data to \textsc{TALYS}
calculations such as those described in the last section can help constrain $%
\gamma $-ray-strength-function models through the $G_{norm}$ factor in Eq. %
\ref{TALYS Eq 4.68}; $G_{norm}>1$ indicates that the $\gamma $-ray strength
function (below $S_{n}$) is too small, and vice versa. However, the
level-density model also enters this equation. Therefore, we calculated the
average $G_{norm}$ for 12 calculations, run for the four Pt isotopes with
the three level-density models (which can be adjusted to measured $D_{0}$
values as described above), for each of the four $\gamma $%
-ray-strength-function models in \textsc{TALYS}. The resulting average $%
G_{norm}$ values are 1.50$\pm $0.80, 0.39$\pm $0.20, 1.61$\pm $0.70, and 1.08%
$\pm $0.46 for models 1--4, respectively, where listed uncertainties are
standard deviations of the distributions. These results indicate that model
4 (Hartree-Fock-Bogolyubov \cite{Kh2005}) yields results closest to the
data, with nearly equal distribution of $G_{norm}$ values below (7/12) and
above (5/12) 1.0. Model 2 (Brink-Axel Lorentzian \cite{Br57,Ax62}) gave the
worst results, with all 12 $G_{norm}$ values substantially smaller than 1.0,
indicating that this $\gamma $-ray-strength-function model is consistently
too large. Models 1 (Kopecky-Uhl generalized Lorentzian \cite{Ko90a}) and 3
(Hartree-Fock BCS \cite{Kh2005}) gave intermediate results with both having $%
G_{norm}$ values larger than 1.0 ($\gamma $-ray-strength function too small)
for 3/4 of the cases. Level-density models 1 (constant temperature plus
Fermi gas \cite{Gi65}) and 2 (back-shifted Fermi gas \cite{Di73})
consistently gave the largest and smallest, respectively $G_{norm}$ values.

\section{Conclusions\label{ConclusionsSection}}

Our new $^{192,194,195,196}$Pt($n,\gamma $) and total cross section data
respresent very large improvements over previous work. Astrophysical ($%
n,\gamma $) reaction rates calculated from our data are substantially
different from, and much more accurate than, currently recommended rates. We
recommend a substantially larger $^{192}$Ir($n,\gamma $) reaction rate with
steeper energy dependence, based in large part on calibration to local
systematics deduced from our new Pt data. In addition, statistical-model
calculations undertaken to assist this effort indicate that neutron strength
functions may be more important for accurately predicting reaction rates for
unmeasured nuclides than is routinely assumed.

\begin{acknowledgments}
This work was supported by the Research Council of Norway and by the Office
of Nuclear Physics of the U.S. Department of Energy under Contract No.
DE-AC05-00OR22725 with UT-Battelle, LLC.
\end{acknowledgments}

\bibliographystyle{prsty}
\bibliography{ACOMPAT,pauls}

\newif\ifabfull\abfulltrue

\end{document}